\documentclass[a4paper,11pt]{article}
\usepackage{pos}

\usepackage{enumitem}

\usepackage{dsfont}
\usepackage{mathrsfs}

\newcommand{\id}[0]{\mathds{1}}
\usepackage{placeins}

\def\d{\partial}
\def\del{\nabla}

\def\a{\alpha}

\usepackage{bm}

\usepackage{caption}
\usepackage{subcaption}

\def\calC{\mathcal{C}}

\def\t{\tau}

\def\Om{\Omega}

\def\l{\lambda}

\def\calI{\mathcal{I}}

\def\ccccend{\end{array}\right)}

\def\t{\tau}

\newcommand{\calK}{\mathcal{K}}

\renewcommand\Re{\operatorname{Re}}

\def\de{\delta}
\def\De{\Delta}

\usepackage[bottom]{footmisc}

\theoremstyle{definition}

\usepackage{amsthm}
\theoremstyle{theorem}

\def\del{\partial}

\def\bbZ{\mathbb{Z}}

\def\calV{\mathcal{V}}

\def\Ombar{\bar{\Om}}

\newcommand{\kom}{\, ,\;}

\def\calK{\mathcal{K}}
\def\calV{\mathcal{V}}
\def\calC{{\mathcal{C}_{\mathrm{cf}}}}

\def\Xt{\widetilde{X}}
\def\Dec{\De^\circ}

\def\I{i}
\def\calF{\mathcal{F}}
\def\calA{\mathcal{A}}

\title{Brief overview of Candidate de Sitter Vacua}
\author*[a,b]{Andreas Schachner}
		
\affiliation[a]{Arnold Sommerfeld Center for Theoretical Physics, Ludwig-Maximilian-University Munich,\\
  Theresienstr. 37, 80333 Munich, Germany}

\affiliation[b]{Department of Physics, Cornell University,Ithaca, NY 14853 USA}

\emailAdd{a.schachner@lmu.de}

\abstract{

We review compactifications of type IIB string theory which produce de Sitter vacua to leading order in the $\alpha^\prime$ and $g_s$ expansions in line with the scenario proposed by Kachru, Kallosh, Linde, and Trivedi. We detail specific Calabi-Yau orientifold compactifications incorporating the non-perturbative superpotential from Euclidean D3-branes, the full flux-induced superpotential, and the Kähler potential evaluated at string tree level but retaining all orders in $\alpha'$. Each model hosts a Klebanov-Strassler throat featuring a single anti-D3-brane. The energy associated with this supersymmetry-breaking source --- computed at leading order in $\alpha'$ --- lifts the minimum to a metastable de Sitter vacuum with all moduli stabilised. A key open challenge is the identification of vacua that remain stable when including additional corrections --- an endeavour for which this study provides a solid foundation. This work is a contribution to the proceedings of the Corfu Summer Institute 2024 "School and Workshops on Elementary Particle Physics and Gravity" (CORFU2024) and is based on \cite{McAllister:2024lnt}.

}

\FullConference{Proceedings of the Corfu Summer Institute 2024 "School and Workshops on Elementary Particle Physics and Gravity" (CORFU2024)\
12 - 26 May, and 25 August - 27 September, 2024\\
Corfu, Greece\\}

 \tableofcontents

\begin{document}
\maketitle

\section{Introduction}\label{sec:intro}

The expansion of the Universe is accelerating. The most basic cosmological model consistent with this observation is de Sitter space.
To understand how gravity might be quantised in our Universe and to confront the cosmological constant problem, it is therefore essential to explore de Sitter vacua within the framework of string theory. Despite the profound significance of this issue, and the vast number of studies on string compactifications (see, for instance, \cite{Candelas:1985en}), explicit examples of de Sitter vacua have proven remarkably difficult to realise.

In this note, we summarise new progress in realising de Sitter vacua through Calabi-Yau compactifications of type IIB string theory \cite{McAllister:2024lnt}. Our approach closely follows the framework proposed over two decades ago by Kachru, Kallosh, Linde, and Trivedi (KKLT) \cite{Kachru:2003aw}, and we assemble for the first time all the essential elements of the KKLT construction within fully explicit models.
Specifically, to realise a KKLT de Sitter vacuum, one must identify a Calabi-Yau orientifold in which the following ingredients can be simultaneously implemented:
\vspace{-0.05cm}
\begin{enumerate}[label=(\alph*)] \item\label{it:req1} a supersymmetric AdS vacuum with an exponentially suppressed vacuum energy,
\vspace{-0.05cm}
\item\label{it:req2} a conifold region that supports a Klebanov-Strassler throat, with a redshifted energy scale comparable to the AdS vacuum energy, and
\vspace{-0.05cm}
\item\label{it:req3} an anti-D3-brane whose contribution uplifts the AdS vacuum to positive vacuum energy.
\end{enumerate}
In this work, we successfully realise conditions (a) through (c) within explicit flux compactifications on Calabi-Yau orientifolds. These constructions are carried out at leading order in the effective field theories (EFTs) describing the compactifications. As such, we provide concrete examples of KKLT-type de Sitter vacua at this level of approximation. This marks the first instance in which such vacua have been explicitly constructed.

We begin by identifying suitable Calabi-Yau orientifolds and choosing quantised flux configurations from which we derive the leading-order effective theories defined precisely in the sections that follow. 
In particular, we incorporate at string tree level all (non-)perturbative corrections in $\alpha'$ to the Kähler potential and the holomorphic Kähler coordinates. Consequently, our treatment of the supersymmetric EFT --- namely, the theory prior to the inclusion of anti-D3-branes --- is exact in $\alpha'$, though not in the string coupling $g_s$.

\begin{table}[t!]
\begin{centering}
\resizebox{\linewidth}{!}{
\begin{tabular}{|c|c|c|c|c|c|c|c|c|c|}
\hline
& & & & & & & & &\\[-1.3em]
ID & $h^{2,1}$ & $h^{1,1}$ & $M$ & $K'$ & $g_s$ & $W_0$ & $g_sM$ & $|z_{\mathrm{cf}}|$  & $V_0$ \\[0.1em]
\hline
\hline
& & & & & & & & &\\[-1.3em]
1 & 8 & 150 & 16 & $\frac{26}{5}$ & 0.0657 & 0.0115 & 1.051 &  2.822$\times10^{-8}$ &  +1.937$\times10^{-19}$ \\[0.1em]\hline 
2 & 8 & 150 & 16 & $\frac{93}{19}$ & 0.0571 & 0.00490 & 0.913 &  7.934$\times10^{-9}$ &  +1.692$\times10^{-20}$ \\[0.1em]\hline
3 & 8 & 150 & 18 & $\frac{40}{11}$ & 0.0442 & 0.0222 & 0.796 &  8.730$\times10^{-8}$ &  +4.983$\times10^{-19}$ \\[0.1em]\hline
4 & 5 & 93 & 20 & $\frac{17}{5}$ & 0.0404 & 0.0539 & 0.808 & 1.965$\times10^{-6}$ &  +2.341$\times10^{-15}$ \\[0.1em]\hline 
5 & 5 & 93 & 16 & $\frac{29}{10}$ & 0.0466 & 0.0304 & 0.746 & 8.703$\times10^{-7}$ &  +2.113$\times10^{-15}$ \\[0.1em]\hline 
\end{tabular}
}
\caption{Five candidate de Sitter vacua together with their control parameters. 
Examples 1 and 4 will be discussed in details in \S\ref{sec:examples}.}
\label{tab:summary}
\end{centering}
\end{table}

\begin{figure}[t!]
\begin{center} 
\includegraphics[width=1\linewidth]{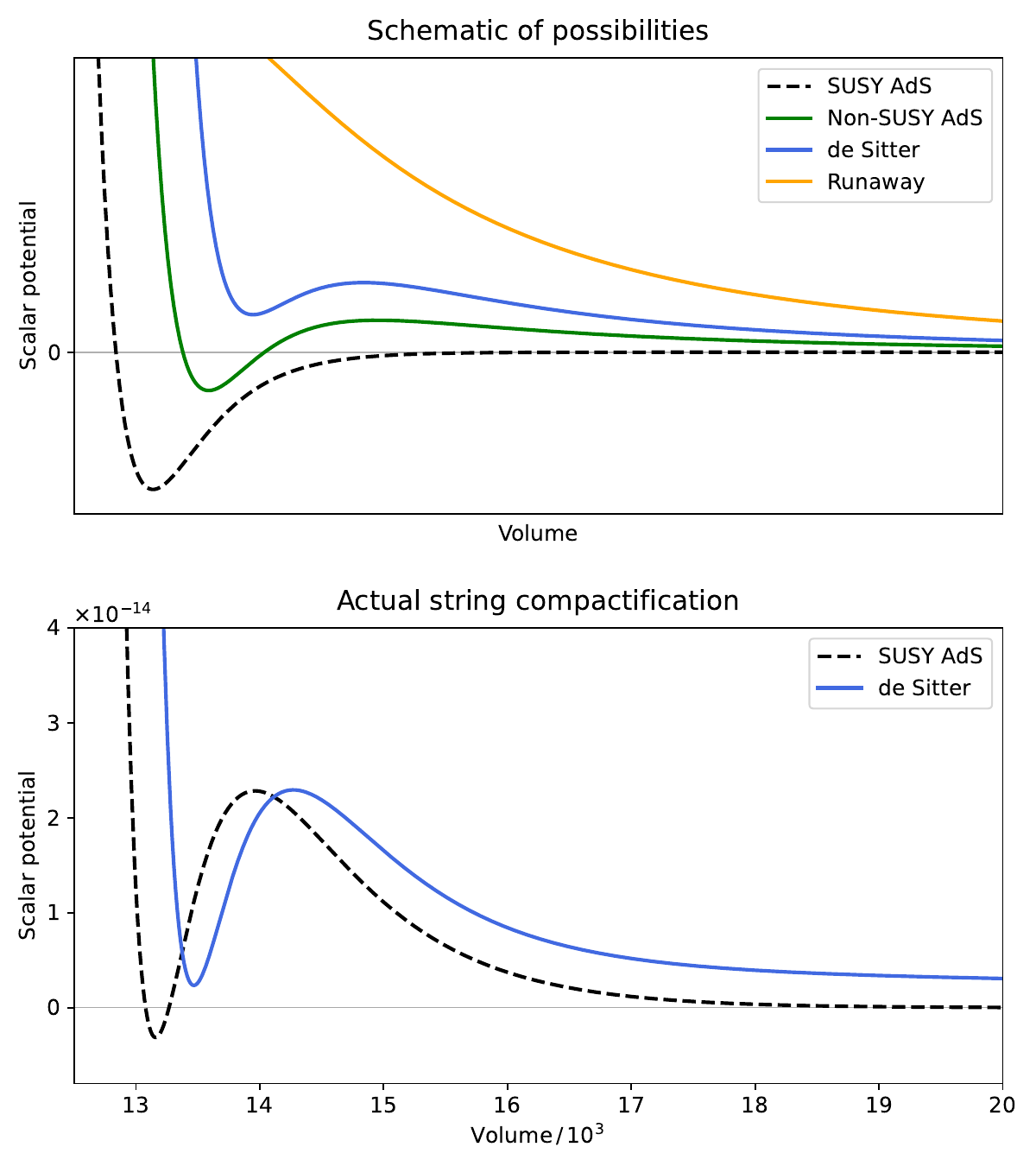}
\caption{
\emph{Top:} Schematic illustration of a supersymmetric AdS vacuum and the possible outcomes of the anti-D3-brane uplift in the KKLT scenario.  
\emph{Bottom:} Scalar potentials corresponding to the AdS vacuum (black) and the uplifted dS vacuum (blue) in the flux compactification on the Calabi-Yau orientifold described in \S\ref{sec:aule}.  
The lower panel presents results from a fully explicit computation within the leading-order EFT defined in \S\ref{sec:EFTs} and~\ref{sec:methods}.  
The horizontal axis denotes the Einstein frame volume $\mathcal{V}_E$ of the Calabi-Yau in units of the string length, while the vertical axis represents the scalar potential $V$ in Planck units.  
Definitions of these quantities are given in Eqs.~\eqref{eq:Vst},
\eqref{eq:Kahlerpotential} and~\eqref{eq:Vfull}.  
The lower panel is reproduced from Fig.~\ref{fig:aule_uplift}.
}\label{fig:introExamplePlot}
\end{center}
\end{figure}

Within this framework, we construct 33{,}371 distinct compactifications, each featuring a Klebanov-Strassler throat containing a single anti-D3-brane. Among these, we identify five configurations in which the anti-D3-brane’s supersymmetry-breaking energy --- computed at leading order in $\alpha'$ --- uplifts the vacuum to a metastable de Sitter state with complete moduli stabilisation which we call \textit{de Sitter vacua at leading order}.
The parameters of the examples are listed in Tab.~\ref{tab:summary}.
We show the scalar potential for Example 4 in Fig.~\ref{fig:introExamplePlot} where the AdS vacuum before uplifting corresponds to the dashed, black line and the uplifted dS vacuum to the blue line.

While higher-order corrections in $\alpha'$ and $g_s$ could possibly have a significant impact, their full structure is not yet understood. Furthermore, there remains ambiguity regarding whether flux quantisation conditions in Calabi-Yau orientifolds allow for odd integer fluxes. As such, our results stop short of establishing a definitive proof that KKLT-type de Sitter vacua exist as solutions within string theory. Nevertheless, the work represents a meaningful step forward in the practical realisation of de Sitter vacua in type IIB compactifications and lays essential groundwork for future developments in this area.

\newpage

\section{Leading-Order Supersymmetric EFT} 
\label{sec:EFTs}

Let us start by introducing the leading-order supersymmetric EFT
in which we construct all our vacua.
In this note, we will only summarise the most relevant aspects of the construction and we refer to \cite{McAllister:2024lnt} for further details.
Unless stated otherwise, we work in 10D Einstein frame in units $\ell_s^2\equiv (2\pi)^2\alpha'=1$, and follow the conventions of \cite{Demirtas:2019sip}.

We consider type IIB string theory compactified on a Calabi-Yau threefold $X$ to four dimensions. 
By imposing an O3/O7 orientifold action \cite{Sagnotti:1987tw,Bianchi:1990yu,Dabholkar:1996pc,Sen:1996vd,Dabholkar:1997zd,Acharya:2002ag,Brunner:2003zm,Brunner:2004zd}, half the supersymmetries get broken leading to a $\mathcal{N}=1$ supergravity theory in four dimensions.
In this theory, the light scalar degrees of freedom from the closed string sector which we denote $\Phi^I$ reside in chiral multiplets.
In all of our examples,  
we work exclusively with involutions $\mathcal{I}:\, X\rightarrow X$ satisfying $h^{1,1}_-(X,{\mathcal{I}})=h^{2,1}_+(X,{\mathcal{I}})=0$
such that the chiral multiplets surviving the orientifold projection are \cite{Grimm:2004uq}
\begin{enumerate}
    \item $h^{1,1}$ complexified \emph{Kähler moduli} $T_i$\,,
    \item $h^{2,1}$ complex structure moduli $z^a$\,, and
    \item the axio-dilaton $\tau\coloneqq C_0+\I/g_s$\,.
\end{enumerate}
The $F$-term potential for these fields
\begin{equation}
    V_F(\Phi,\overline{\Phi}) = \mathrm{e}^{\mathcal{K}}\left(\calK^{I\bar{J}}D_IWD_{\bar{J}}\bar{W} - 3\left|W\right|^2\right)\,,\label{eq:VF}
\end{equation}
is fully specified by a non-holomorphic Kähler potential $\calK$
and a holomorphic superpotential $W$.
Here, we introduced the covariant derivative 
\begin{equation}
    D_IW = \d_IW + \d_I\calK \,W\,,\label{eq:DWdef}
\end{equation}
and the Kähler metric 
\begin{equation}
    \calK_{I\bar{J}} = \d_I\d_{\bar{J}}\calK\,.
\end{equation} 
Accordingly, the evaluation of $V_F$ requires the specification of the Kähler potential $\calK$ and superpotential $W$, together with an explicit parametrisation of the holomorphic fields $\Phi^I$ in terms of the underlying compactification data.

In the absence of a full understanding of quantum corrections, we need to make suitable assumptions and truncations which leads us to defining a \emph{leading-order supersymmetric EFT}.
That is, we will work in a leading-order approximation for the Kähler potential $\calK \approx \calK_{\text{l.o.}}$ (see Eq.~\eqref{eq:detailedform2}) and the Kähler coordinates $T_i \approx T^{\text{l.o.}}_i$ (see Eq.~\eqref{eq:detailedform4}) incorporating all the known $\alpha'$ corrections at string tree level.
Together with the superpotential $W$ defined in Eq.~\eqref{eq:wlo3},
this allows us to specify an $\mathcal{N}=1$ supersymmetric supergravity theory
with the $F$-term potential \eqref{eq:VF} corresponding to
\begin{equation}\label{eq:vfsum}
    V_F = V_F(W;\mathcal{K}_{\text{l.o}};\tau, z^a, T^{\text{l.o.}}_i)\,.
\end{equation}
This potential specifies the dynamics for all moduli in terms of topological data for mirror pairs of
Calabi-Yau threefolds $(X,\widetilde{X})$,
orientifold involutions $\mathcal{I}:\, X\rightarrow X$
and quantised fluxes, cf.~\S\ref{sec:methods}.
In addition to \eqref{eq:vfsum},
in the presence of an anti-D3-brane we have to add the uplifting potential which we will take as the
leading-order contribution $V_{\overline{D3}}$ in the $\alpha'$ expansion \cite{Kachru:2002gs} as specified in Eq.~\eqref{eq:anti-D3-potential0}.

\newpage

\subsection{The Kähler potential}

The Kähler potential $\calK $ that will be used to evaluate the $F$-term potential \eqref{eq:VF} can be written as
\begin{align}
\calK \approx \calK_{\text{l.o.}} \coloneqq  &\,\,\calK_{\text{tree}} + \calK_{(\alpha')^3} + \calK_{\text{WSI}}\, \\
= & -2\log\Bigl(2^{3/2}g_s^{-3/2}\calV \Bigr) -\log\bigl(-\I\left(\t-\bar{\t}\right)\bigr) - \log\Bigl(-\I\int_{X}\Om\wedge\Ombar\Bigr)\,, \label{eq:detailedform2}
\end{align}
in terms of the holomorphic $3$-form $\Omega(z^{a})$ (cf.~Eq.~\eqref{eq:PiVec}), and
the string tree level, $\alpha'$-corrected, string-frame volume $\calV$ (cf.~Eq.~\eqref{eq:Kahlerpotential}).

\subsubsection*{Kähler moduli sector}

Initially, we introduce a basis  $\{\omega^i\}_{i=1}^{h^{1,1}(X)}$
of $H^4(X,\mathbb{Z})$ and its dual basis $\{\omega_i\}_{i=1}^{h^{1,1}(X)}$ of $H^2(X,\mathbb{Z})$ satisfying $\int_X\omega^i\wedge  \omega_j={\delta^i}_j$.
The Kähler cone $\mathcal{K}_X\subset H^{1,1}(X,\mathbb{R})$ of $X$ is parametrised by the Kähler parameters $\{t^i\}_{i=1}^{h^{1,1}(X)}$.
With these definitions at hand,
we can write the string-frame Kähler class $J$ and the triple intersection numbers $\kappa_{ijk}$ of $X$ as
\begin{equation}
J=\sum_i t^i\,\omega_i\, ,\; \kappa_{ijk}\coloneqq \int_X \omega_i\wedge \omega_j\wedge \omega_k\, .
\end{equation}
The cone dual to $\mathcal{K}_X$ is 
The Mori cone $\mathcal{M}_{X} \subset H_{2}(X,\mathbb{R})$ of $X$ is the cone dual to the Kähler cone $\mathcal{K}_X$
and thus $q_it^i>0$ for all $\vec{t}\in\mathcal{K}_X$ for any $\mathbf{q}\in\mathcal{M}_{X}$.

The classical string-frame volume $\calV^{(0)}$ and Einstein-frame volume $\calV_E^{(0)}$
of $X$ can be expressed in terms of the string-frame Kähler parameters $t^i$ and the triple intersection numbers $\kappa_{ijk}$ of $X$ as
\begin{equation}\label{eq:Vst}
    \calV^{(0)} = \frac{1}{6}\kappa_{ijk}t_it_jt_k\, ,\; \calV_E^{(0)} = \frac{\calV^{(0)}}{g_s^{3/2}}\, .
\end{equation}
The corrected string-frame volume $\calV$ appearing in \eqref{eq:detailedform2} reads
\begin{align}\label{eq:Kahlerpotential}
\mathcal{V} &=\calV^{(0)}+\delta\calV_{(\alpha')^3} +\delta\calV_{\text{WSI}}\,,
\end{align}
in terms of the tree level $(\alpha')^3$ correction \cite{Grisaru:1986kw,Gross:1986iv,Antoniadis:1997eg,Becker:2002nn}
\begin{align}\label{eq:Valphap3}
    \delta\calV_{(\alpha')^3} &= -\frac{\zeta(3)\chi(X)}{4(2\pi)^3}\, , 
\end{align}
and worldsheet instanton corrections \cite{Dine:1986zy,Dine:1987bq,Grimm:2007xm}
\begin{equation}\label{eq:VWSI}
     \delta\calV_{\text{WSI}} = \frac{1}{2(2\pi)^3}
\sum_{\mathbf{q}\in \mathcal{M}_{X}}\, \mathscr{N}_{\mathbf{q}}\,\Biggl( \text{Li}_3\Bigl((-1)^{\mathbf{\gamma}\cdot \mathbf{q}}\mathrm{e}^{-2\pi \mathbf{q}\cdot \mathbf{t}}\Bigr) + 2\pi \mathbf{q}\cdot \mathbf{t}\,\,\text{Li}_2\Bigl((-1)^{\mathbf{\gamma}\cdot \mathbf{q}}\mathrm{e}^{-2\pi \mathbf{q} \cdot \mathbf{t}}\Bigr)\Biggr)\,,
\end{equation}
where $\mathscr{N}_{\mathbf{q}}$ are genus-zero Gopakumar-Vafa (GV) invariants \cite{Gopakumar:1998ii,Gopakumar:1998jq} of $X$ and $\gamma^i\coloneqq \int_X [\text{O7}]\wedge \omega^i$.
Further, we note that polylogarithms $\text{Li}_k(z)$ are defined for $|z|<1$ as 
\begin{equation}\label{eq:polylog}
    \text{Li}_k(z)=\sum_{n=1}^\infty \frac{z^n}{n^k}\, ,
\end{equation}
and can be analytically continued to the entire complex plane.

Next, let us define the Kähler moduli $T_i$ corresponding to be the natural Kähler coordinates to compute the Kähler metric $\calK_{I\bar{J}}$.
At leading order, there is a simple relationship between the holomorphic coordinates $T_i$ and the curve volumes $t^i$
which is modified in the presence of perturbative corrections to $\calK_{I\bar{J}}$.
Indeed, the Kähler coordinates can be written as (see e.g.~\cite{Cecotti:1988qn,Grimm:2004uq,Robles-Llana:2006hby,Robles-Llana:2007bbv,Grimm:2007xm,Baume:2019sry,Marchesano:2019ifh})
\begin{align}\label{eq:detailedform4}
T_i \approx T^{\text{l.o.}}_i
\coloneqq \frac{1}{g_s}\Bigl(\mathcal{T}^{\text{tree}}_i+ 
\mathcal{T}_i^{(\alpha')^2}    
+
\mathcal{T}_i^{\text{WSI}}\Bigr) +i\int_X C_4\wedge \omega_i\,,
\end{align}
with 
%\begin{align}
%\mathcal{T}^{\text{tree}}_i+ 
%\mathcal{T}_i^{(\alpha')^2}   
%+
%\mathcal{T}_i^{\text{WSI}} 
%&=\frac{1}{2}\kappa_{ijk}t^jt^k-\frac{\chi(D_i)}{24}\nonumber\\
%&\hphantom{=}+\frac{1}{(2\pi)^2}\sum_{\mathbf{q}\in \mathcal{M}_{X}}q_i\, \mathscr{N}_{\mathbf{q}} \,\text{Li}_2\Bigl((-1)^{\mathbf{\gamma}\cdot \mathbf{q}}\mathrm{e}^{-2\pi \mathbf{q}\cdot \mathbf{t}}\Bigr)\,. 
%\end{align}
\begin{align}
\mathcal{T}^{\text{tree}}_i &= \frac{1}{2}\kappa_{ijk}t^jt^k \, ,\\[0.3em]
\mathcal{T}_i^{(\alpha')^2}   &= \frac{\chi(D_i)}{24} \, ,\\[0.3em]
\mathcal{T}_i^{\text{WSI}} &=\frac{1}{(2\pi)^2}\sum_{\mathbf{q}\in \mathcal{M}_{X}}q_i\, \mathscr{N}_{\mathbf{q}} \,\text{Li}_2\Bigl((-1)^{\mathbf{\gamma}\cdot \mathbf{q}}\mathrm{e}^{-2\pi \mathbf{q}\cdot \mathbf{t}}\Bigr)\,. 
\end{align}

\subsubsection*{Complex structure sector}

Coming back to \eqref{eq:detailedform2}, we would like to compute the third term depending on the complex structure moduli $z^{a}$ through the holomorphic $3$-form $\Omega(z^{a})$.
To begin with, we rewrite the integral over $\Omega(z^{a})$ in terms of a \emph{period vector} $\vec{\Pi}(z^{a})$ as
\begin{equation}
    \int_{X}\Om\wedge\Ombar = \vec{\Pi}^\dagger\cdot\Sigma\cdot\vec{\Pi}\; , \quad \Sigma \coloneqq  \begin{pmatrix} 0 & \id\\
                -\id & 0
                \end{pmatrix}\,.
\end{equation}
By working in a symplectic basis of $H_3(X,\mathbb{Z})$ and with an appropriate normalisation of $\Omega$,
the period vector $\vec{\Pi}$ can be written in terms of a holomorphic prepotential $\mathcal{F}(z^{a})$ as
\begin{equation}\label{eq:PiVec} 
   \vec{\Pi}=\left ( \begin{array}{c}
        2\mathcal{F}-z^a\del_{z^a}\mathcal{F}  \\
        \partial_{z^a} \mathcal{F}(z)  \\
        1  \\
        z^a
   \end{array}\right ) \, .
\end{equation}
Here,
the $z^a$, $a=1,\ldots,h^{2,1}(X)$, act as local affine coordinates on complex structure moduli space $\mathcal{M}_{\text{cs}}(X)$ of $X$.

The prepotential $\mathcal{F}(z^{a})$ can be computed explicitly in the \emph{large complex structure} (LCS) patch.
Indeed, at LCS the prepotential can be expanded in terms of the topological data of the mirror Calabi-Yau threefold $\Xt$ as the sum of a polynomial piece and an exponential piece \cite{Hosono:1994av}
\begin{align}\label{eq:prepotential}
\mathcal{F}(z^{a})=\mathcal{F}_{\text{poly}}(z^{a})+\mathcal{F}_{\text{inst}}(z^{a})\,,
\end{align}
where
\begin{align}
    \label{eq:Fpoly}\mathcal{F}_{\text{poly}}(z^{a})&=-\frac{1}{3!}\widetilde{\kappa}_{abc}z^az^bz^c+\frac{1}{2}a_{ab}z^az^b+\frac{1}{24}\tilde{c}_a z^a+\frac{\zeta(3)\chi(\widetilde{X})}{2(2\pi \I)^3}\,
    ,\\[0.5em]
    \label{eq:Finst}\mathcal{F}_{\text{inst}}(z^{a})&=-\frac{1}{(2\pi \I)^3}\sum_{\tilde{\mathbf{q}}\in \mathcal{M}_{\Xt}}\mathscr{N}_{\tilde{\mathbf{q}}}\,\text{Li}_3\Bigl(\mathrm{e}^{2\pi \I\,\tilde{\mathbf{q}}\cdot \mathbf{z}}\Bigr)\, .
\end{align}
In \eqref{eq:Fpoly}, $\widetilde{\kappa}_{abc}$ denotes the triple intersection numbers of $\widetilde{X}$, while
\begin{equation}
\tilde{c}_a=\int_{\widetilde{X}}c_2(\widetilde{X})\wedge \tilde{\beta}_a\, ,\quad
a_{ab}\equiv\dfrac{1}{2} \begin{cases}
\widetilde{\kappa}_{aab} & a\geq b\\
\widetilde{\kappa}_{abb} & a<b
\end{cases}\, , \quad \text{and} \quad \chi(\widetilde{X})=\int_{\widetilde{X}} c_3(\widetilde{X})\, . \label{eq:amatrix}
\end{equation}
in terms of a basis $\{\tilde{\beta}_a\}_{a=1}^{h^{2,1}(X)}$ of $H^2(\widetilde{X},\mathbb{Z})$, and $c_n(\widetilde{X})$ denotes the $n$-th Chern class of $\widetilde{X}$. 
The instantonic piece $\calF_{\text{inst}}$ in \eqref{eq:Finst} involves a sum over effective curve classes $\tilde{\mathbf{q}}$ in $H^4(\widetilde{X},\mathbb{Z})\simeq H_2(\widetilde{X},\mathbb{Z})$ where
the coefficients $\mathscr{N}_{\tilde{\mathbf{q}}}$ are the genus-zero Gopakumar-Vafa (GV) invariants \cite{Gopakumar:1998ii,Gopakumar:1998jq} of $\widetilde{X}$.

\subsubsection*{Engineering conifolds}

To later construct a Klebanov-Strassler throat region in a flux compactification, we must stabilise the complex structure moduli near a \emph{conifold singularity}.
In the complex structure moduli space $\mathcal{M}_{\text{cs}}(X)$ of $X$,
such a conifold locus arises whenever a set of $n_{\mathrm{cf}}$ 3-cycles in some homology class $[\mathcal{C}] \in H_3(X,\mathbb{Z})$ shrink to zero volume. 
Because $\mathcal{M}_{\text{cs}}(X)$ is identified with the complexified  Kähler cone $\calK_{\Xt}$ of the mirror threefold $\Xt$ in an LCS patch, the conifold locus corresponds to a facet $\mathcal{K}_{\mathrm{cf}}$ of 
$\calK_{\Xt}$.\footnote{Strictly speaking, we denote by $\mathcal{K}_{\mathrm{cf}}$ the \emph{interior} of said facet where the set conifold curves shrinks.}
The set of curves $\calC$ shrinking at this facet are in some effective curve class  $\tilde{\mathbf{q}}_{\mathrm{cf}} \in \mathcal{M}_{\Xt}\cap H_2(\Xt,\mathbb{Z})$ which we refer to as the \textit{conifold class}, see e.g. \cite{Demirtas:2020ffz}. 
Assuming that such as conifold class $\tilde{\mathbf{q}}_{\mathrm{cf}}$ exists,
the volume of the curves $\calC$ is measured by the absolute value of
\emph{conifold modulus} $z_{\text{cf}}$.
For convenience, we choose a basis for $H_2(\Xt,\mathbb{Z})$ in which the conifold curve is represented by $\tilde{\mathbf{q}}_{\mathrm{cf}}=(1,0,\ldots,0)$ and hence the conifold modulus $z_{\text{cf}}$ is identified as $z_{\text{cf}}  = z^{1}$.
The remaining \emph{bulk moduli} are then denoted as $z^\alpha$, $\alpha = 2,\ldots,h^{2,1}$.

In the presence of a conifold singularity, we need to analytically continue \eqref{eq:Finst} in the presence of a shrinking curve parametrised by the absolute value of $\tilde{\mathbf{q}}_{\text{cf}}\cdot \mathbf{z}=z_{\text{cf}}$.
This is achieved by making use of Euler’s reflection formula \cite{Demirtas:2020ffz}
\begin{equation}
	-\dfrac{\text{Li}_3\Bigl(\mathrm{e}^{2\pi \I\, z_{\text{cf}}}\Bigr)}{(2\pi \I)^{3}} =  \dfrac{z_{\text{cf}}^{2}}{4\pi \I}\,\ln(-2\pi\I z_{\text{cf}}) - \dfrac{1}{(2\pi \I)^{3}}\sum_{n=0}^{\infty}\, \dfrac{\hat{\zeta}(n-3)}{n!}\,  (2\pi \I z_{\text{cf}})^{n}
\end{equation}
with $\hat{\zeta}(x) = \zeta(x)$ for $x\neq 1$ and $\hat{\zeta}(1)=3/2$.
We can then systematically expand the full prepotential \eqref{eq:prepotential} around small $|z_{\text{cf}}|\ll 1$ which leads to \cite{Demirtas:2020ffz}
\begin{equation}\label{eq:FconiLCS} 
	\mathcal{F}(z_{\text{cf}},z^{\alpha}) = n_{\text{cf}}\, \dfrac{z_{\text{cf}}^{2}}{4\pi \I}\,\ln(-2\pi\I z_{\text{cf}})+\sum_{n=0}^{\infty}\, \dfrac{\mathcal{F}^{(n)}(z^\alpha)}{n!}\,  z_{\text{cf}}^{n}
\end{equation} 
where $n_{\text{cf}}=\mathscr{N}_{\tilde{\mathbf{q}}_{\text{cf}}}$ and
\begin{equation}
\mathcal{F}^{(n)}(z^\alpha) = (\partial_{z_{\text{cf}}}^n \mathcal{F}_{\mathrm{poly}})\bigl |_{z_{\text{cf}}=0} - n_{\text{cf}}\, \dfrac{\hat{\zeta}(3-n)}{(2\pi\mathrm{i})^{3-n}}\,- \dfrac{1}{(2\pi\mathrm{i})^{3-n}}\, \sum_{\tilde{\mathbf{q}}\neq \tilde{\mathbf{q}}_{\mathrm{cf}}}\,  \mathscr{N}_{\tilde{\mathbf{q}}}\, (\tilde{q}_{1})^n\, \mathrm{Li}_{3-n}\Bigl(\mathrm{e}^{2\pi \I\,\tilde{\mathbf{q}}\cdot \mathbf{z}}\Bigr)\bigl |_{z_{\text{cf}}=0}
\end{equation}
in terms of $\mathcal{F}_{\mathrm{poly}}$ as defined in \eqref{eq:Fpoly}.
This expression can be used to compute the periods to any desired order in the expansion of the control parameter $|z_{\text{cf}}|$.
In many cases, expanding to linear order in $z_{\text{cf}}$ is sufficient, though ultimately we work with the full theory in our numerical search for flux vacua described in \S\ref{sec:fluxes}.

\subsection{The superpotential}

Next, let us look at the superpotential which receives two contributions from fluxes and non-perturbative effects.
It can be written as
\begin{align}
W &= W_{\text{flux}} + W_{\text{np}}\,  .\label{eq:wlo2} 
\end{align}
The flux superpotential $W_{\text{flux}}$ encodes the contributions of the background fluxes to the scalar potential \cite{Gukov:1999ya}. 
It can be written as \cite{Gukov:1999ya,Giddings:2001yu},
\begin{align}\label{eq:flux_superpotential}
     W_{\text{flux}}(\tau,z^a)=\sqrt{\tfrac{2}{\pi}}\int_X (F_3-\tau H_3)\wedge \Omega(z)=\sqrt{\tfrac{2}{\pi}}\,\vec{\Pi}^\top \,{\cdot}\,\Sigma\,{\cdot}\, (\vec{f}-\tau \vec{h})\, ,
\end{align}
where $F_3,H_3$ are the RR and NSNS $3$-forms respectively and $\vec{f}$, $\vec{h}\in H^3(X,\bbZ)$ the corresponding flux vectors.
These flux vectors are constrained by Gauss's law
\begin{equation}
     2\left(N_{\text{D3}}-N_{\overline{\text{D3}}}\right) + Q_{\text{flux}} - Q_{\text{O}} = 0\; ,\quad  Q_{\text{flux}} \coloneqq  \int_X H_3 \wedge F_3 = \vec{f}\,^\top \Sigma \vec{h}\; ,\quad  Q_{\text{O}} \coloneqq  \frac{1}{2}\chi_f \,,\label{eq:gaussLaw}
\end{equation} 
in terms of the number of spacetime-filling (anti-)D3-branes $N_{\text{D3}}$ ($N_{\overline{\text{D3}}}$) in the system
and where $\chi_f$ is the Euler character of the fixed locus of $\mathcal{I}$ in $X$.
Given a choice of a Calabi-Yau orientifold $X/\calI$ and flux vectors $\vec{f}$, $\vec{h}$, we will compute the explicit expression for the flux superpotential in \S\ref{sec:fluxes} by using the period vector $\vec{\Pi}$ in Eq.~\eqref{eq:PiVec} as described in the previous subsection.

The non-perturbative superpotential $W_{\text{np}}$ can be written as \cite{Witten:1996bn}
\begin{equation}\label{eq:WnpTMP} 
	W_{\text{np}} = \sum_D\,  \calA_D \mathrm{e}^{-\frac{2\pi}{c_D}T_D}
\end{equation}
in terms of Kähler moduli $T_{D}$ associated to divisors $D$ hosting some D-brane configurations and $c_{D}$ the dual Coxeter number of the corresponding gauge theory.
Indeed, $W_{\text{np}}$
summarises the contributions of Euclidean D3-branes and gaugino condensation on seven-branes to the potential \cite{Witten:1996bn}.
If a divisor $D$ is rigid and smooth, then only two universal fermionic zero modes contribute to the path integral and D-branes wrapped on $D$ contribute to the superpotential with non-vanishing Pfaffian $\calA_D$ \cite{Witten:1996bn}, see \cite{Blumenhagen:2009qh,Schachner:2022bak} for reviews.
Further, for a \emph{pure rigid} divisor $D$, i.e., divisors for which the intermediate Jacobian is trivial (see e.g. \cite{Demirtas:2021nlu,Jefferson:2022ssj}),
the Pfaffian prefactors $\calA_D$ are non-zero constants (called \emph{Pfaffian number}) up to warping effects \cite{Ganor:1996pe,Berg:2004ek,Giddings:2005ff,Shiu:2008ry,Baumann:2006th,Frey:2013bha,Martucci:2016pzt}.
In this work we will only incorporate the contributions of \emph{pure rigid prime toric divisors}, cf.~\S\ref{sec:polytopeselection}.
This restriction was validated \emph{a posteriori} in each example in \cite{McAllister:2024lnt}.

Let us make two remarks.
First, in our constructions, we cancel the D7-brane tadpole locally by putting a stack of four D7-branes on top of each O7-plane giving rise to a $\mathfrak{so}(8)$ $\mathcal{N}=1$ super Yang-Mills theory on that four-cycle, see e.g. \cite{Crino:2022zjk} for alternatives.
Any pure rigid prime toric divisor $D$ therefore has $c_D=6$ if $D$ hosts an O7-plane and $c_D=1$ otherwise. 
Second,
it is presently unknown how to compute the Pfaffian numbers $\calA_D$, see however \cite{Kim:2022jvv,Alexandrov:2022mmy,Kim:2022uni,Kim:2023cbh} for recent progress.
Based on \cite{Alexandrov:2022mmy}, it is convenient to adopt the following normalisation
\begin{equation}
    \calA_D = \sqrt{\frac{2}{\pi}}\frac{n_D}{4\pi^2}\,,\label{eq:K0}
\end{equation}
where $n_D$ is a constant that can be expressed as an integral over worldsheet modes \cite{Alexandrov:2022mmy,Kim:2023cbh}.  
For concreteness, we will set $n_{D} = 1$ for all pure rigid divisors $D$ contributing to \eqref{eq:WnpTMP}. 
For the examples presented in  \S\ref{sec:examples}, 
we verified that the de Sitter vacua exist through the full range $10^{-3} \le n_D \le 10^4$.
While it would be highly exciting to compute $\calA_D$ directly,
our ignorance of the actual values of the Pfaffian numbers does not constitute a significant weakness in our setup.

To summarise,
the full superpotential in our leading-order EFT is given by
\begin{align}
W &=\sqrt{\tfrac{2}{\pi}}\,\vec{\Pi}^\top \,{\cdot}\,\Sigma\,{\cdot}\, (\vec{f}-\tau \vec{h}) + \sqrt{\frac{2}{\pi}}\frac{1}{4\pi^2} \sum_D\,  \mathrm{e}^{-\frac{2\pi}{c_D}T_D}\,,
\label{eq:wlo3}    
\end{align}
where $\vec{\Pi}$ is computed from \eqref{eq:PiVec} in terms of the prepotential \eqref{eq:FconiLCS}
and the sum over $D$ runs over all pure rigid prime toric divisors.

\subsection{Towards anti-D3 uplifting}\label{sec:KS} 

This almost completes our definition of the leading-order EFT.
The above data for the Kähler potential \eqref{eq:detailedform2},
the Kähler moduli \eqref{eq:detailedform4} and the superpotential \eqref{eq:wlo3} is sufficient to compute the $F$-term potential \eqref{eq:vfsum} and to find well-controlled supersymmetric AdS$_4$ vacua of type IIB string theory as constructed in \cite{Demirtas:2021nlu}. 
The goal of this work is instead to obtain de Sitter minima in which supersymmetry is broken by anti-D3-branes sitting at the tip of a Klebanov-Strassler (KS) throat.

Warped KS throats emerge provided that complex structure moduli are stabilised sufficiently close to a conifold locus where $\langle|z_{\mathrm{cf}}|\rangle \ll 1$ \cite{Klebanov:2000hb,Giddings:2001yu}. 
Spacetime-filling anti-D3-branes in the infrared region of these throats provide a positive source of energy possibly uplifting AdS$_4$ vacua to de Sitter \cite{Kachru:2003aw}, thereby breaking supersymmetry.
Including $p$ anti-D3-branes, the full scalar potential reads
\begin{equation}\label{eq:Vfull} 
    V = V_F + V^{\text{up}}\; ,\quad  V^{\text{up}} = p V_{\overline{D3}} + \Delta V^{\overline{D3}}_{(\alpha')^2} + \ldots\,,
\end{equation}
in terms of the $F$-term potential $V_F$ defined in \eqref{eq:VF} and the new contribution $V^{\text{up}}$ coming from the anti-D3-branes.
The leading-order potential
$V_{\overline{D3}} $
was derived by Kachru, Pearson, and Verlinde (KPV) \cite{Kachru:2002gs} and employed by KKLT to find dS$_4$ vacua \cite{Kachru:2003aw}.
The potential $V_{\overline{D3}}$ computed in KPV can be expressed in terms of the Einstein-frame volume $\mathcal{V}_E$ of $X$ and string-frame volume $\widetilde{\calV}$ of $\Xt$ as
\cite{Kachru:2002gs,Kachru:2003aw,Kachru:2003sx}
\begin{equation}
\label{eq:anti-D3-potential0}
    V_{\overline{D3}} = \frac{c}{\calV_E^{4/3}}\; ,\quad c = \frac{\eta\,z_{\mathrm{cf}}^{4/3}}{g_sM^2 \widetilde{\calV}^{2/3}}
    \; ,\quad \eta \approx 2.6727\, .
\end{equation}
Sub-leading $(\alpha')^2$ corrections encoded in $ \Delta V^{\overline{D3}}_{(\alpha')^2}$ have been partially computed in \cite{Junghans:2022exo,Junghans:2022kxg,Hebecker:2022zme,Schreyer:2022len,Schreyer:2024pml}, but will be neglected subsequently.

For a configuration with $p$ anti-D3-branes at the bottom of a KS throat to be metastable,
the KPV analysis \cite{Kachru:2002gs} to leading order in the $\alpha'$ expansion leads to the bound ${M}/{p}\gtrsim 12$.
%\begin{equation}\label{eq:KPV_constraint}
%    \dfrac{M}{p}\gtrsim 12\, .
%\end{equation}
As studied in \cite{Junghans:2022exo,Junghans:2022kxg,Hebecker:2022zme,Schreyer:2022len,Schreyer:2024pml}, this constraint is modified in the presence of $\alpha'$ corrections controlled by $1/(g_sM)$.
In the following, we will focus on configurations with $p=1$ and require $M>12$ and $g_s M\gtrsim 1$.

This completes our discussion of the leading-order EFT that we use below to explicitly construct \emph{candidate} de Sitter vacua.
In the next section, we describe in detail the choices of mirror pairs $(X,\Xt)$ 
before describing our numerical search for de Sitter minima in the full potential $V_{F}+V_{\overline{D3}} $ obtained by combining the $F$-term potential \eqref{eq:vfsum} with the KPV potential \eqref{eq:anti-D3-potential0}.

\section{Construction and search procedure}
\label{sec:methods}

Our next task is to find explicit Calabi-Yau compactifications in which all the relevant contributions in our leading-order EFT can be computed.
In this section, we summarise the concrete setup and lay out our search procedure to find suitable candidate geometries for our construction.
 
\subsection{Orientifolds from the Kreuzer-Skarke list}\label{sec:polytopeselection}

Let $\Delta,\Delta^{\circ} \subset \mathbb{Z}^4$ be a pair of polar dual four-dimensional reflexive polytopes from the Kreuzer-Skarke list \cite{Kreuzer:2000xy}.
We restrict to the case where both polytopes are \emph{favourable}, see e.g. \cite{Demirtas:2018akl} for definitions.
A fine, regular, star triangulation (FRST) $\mathscr{T}$ of $\Delta^{\circ}$
defines a toric fan for a four-dimensional toric variety $V$ in which a smooth Calabi-Yau threefold $X$ is realised as the generic anti-canonical hypersurface.
The Cox ring is generated by $h^{1,1}(X)+4$ toric coordinates $x_I$.
Prime toric divisors of $V$ are defined as $\mathscr{D}_I \coloneqq  \{x_I = 0\}$
which descend to prime toric divisors of $X$ through $D_I \coloneqq  \mathscr{D}_I \cap X$ which generate $H_4(X,\mathbb{Z})$.

There are two important requirements on the choice of polytope pairs $(\Delta,\Delta^{\circ})$.
First, in order to apply the mechanism of \cite{Kachru:2003aw} to stabilise all $h^{1,1}(X)$ Kähler moduli $T_i$,
we need to ensure at least $h^{1,1}(X)$ contributions in the non-perturbative superpotential \eqref{eq:wlo3}.
For this reason, we restrict our attention to polytopes $\Delta^{\circ}$ which have at least $h^{1,1}(X)$ rigid prime toric divisors.
Second, the presence of a conifold singularity in $X$ is achieved by shrinking a toric flop curve $\mathcal{C}_{\mathrm{cf}}$ in the mirror $\widetilde{X}$ defined by an FRST $\widetilde{\mathscr{T}}$ of $\Delta$.
In this work, we restrict to polytopes $\Delta$ admitting conifold curves $\mathcal{C}_{\mathrm{cf}}$ with GV invariant $\mathscr{N}_{\mathcal{C}_{\mathrm{cf}}}=2$ implying that there are two conifold singularities.

For such Calabi-Yau threefold hypersurfaces $X$,
the methods of \cite{Moritz:2023jdb} allow us to enumerate those holomorphic, isometric involutions ${\mathcal{I}}:X\rightarrow X$ that are inherited from the ambient variety $V$.
We are specifically interested in orientifolds $X/\calI$ which satisfy $h^{1,1}_-=h^{2,1}_+=0$.
They are obtained from toric varieties $V$ for which the underlying polytope $\Delta^{\circ}$ is \emph{trilayer} \cite{Moritz:2023jdb}.
Indeed, this 
additional polytope property 
ensures the existence of a certain toric coordinate, say $x_1$, for which the involution of $V$ defined by $x_1 \to -x_1$ leads to $h^{1,1}_-=h^{2,1}_+=0$.
In what follows, we refer to such orientifolds as \emph{trilayer orientifolds}.

We make sure that the orientifold symmetry $x_1 \to -x_1$ exchanges the two conifold singularities.
Since there is always a prime toric divisor $D_{\mathrm{cf}}\coloneqq \{x_{\mathrm{cf}}=0\}$ intersecting the two conifolds,
we have to slightly revise our restriction on the number of rigid prime toric divisors from above.
This is because a Euclidean D3-brane on $D_{\mathrm{cf}}$ passes through the highly warped throat region and thus its contribution to the non-perturbative superpotential \eqref{eq:wlo3} will be suppressed exponentially compared to the other terms.
Since this warped Euclidean D3-brane does effectively not contribute to Kähler moduli stabilisation,
we need to ensure that there are at least $h^{1,1}$ rigid prime toric divisors \emph{excluding} $D_{\mathrm{cf}}$ which is necessary for to the existence of a KKLT point.

The D3-brane tadpole \eqref{eq:gaussLaw} for trilayer orientifolds is given by $Q_{\text{O}} =  2+h^{1,1}+h^{2,1}$.
%\begin{equation}\label{eq:qdef}
%Q_{\text{O}} =  2+h^{1,1}+h^{2,1}\, .
%\end{equation}
Typically, the larger $Q_{\text{O}} $, the richer the underlying search space of flux vacua in flux compactifications on $X/\calI$.
Thus, it is beneficial to search for polytopes with sufficiently large Hodge numbers.
However, since the construction of flux vacua described in \S\ref{sec:fluxes} becomes expensive at large $h^{1,2}\gtrsim 10$,
we restrict our search to the range $3 \le h^{2,1} \le 8$.
Even so, the D3-tadpole can be made large $Q_{\text{O}}\ge 100$ provided $h^{1,1}$ is large.
In total, we find 416 Calabi-Yau orientifolds meeting all of the requirements from above, see also Tab.~\ref{tab:scan_summary}.

\subsection{Selection of fluxes and complex structure moduli stabilisation}\label{sec:fluxes}

Given a choice of Calabi-Yau orientifold $X/\calI$ and conifold curve $\calC$,
we now aim at selecting suitable flux vectors $\vec{f},\vec{h}\in H^{3}(X,\bbZ)$ stabilising the complex structure moduli close to the conifold locus, while achieving small $|W_0|$ \cite{Demirtas:2020ffz} and allowing for a single anti-D3-brane by Gauss' law \eqref{eq:gaussLaw}.

Let us evaluate the flux superpotential \eqref{eq:flux_superpotential} and
Kähler potential \eqref{eq:detailedform2} as an expansion in $z_{\mathrm{cf}}$. 
We choose quantised fluxes \cite{Demirtas:2020ffz}
\begin{equation}
\vec{f} = \left(P_0,P_a,0,M^a\right)^\top \, ,\; \vec{h} = \left(0,K_a,0,0^a\right)\,.
\end{equation}
Using \eqref{eq:FconiLCS} at leading order in $z_{\text{cf}}$,
the flux superpotential \eqref{eq:flux_superpotential} reads
\begin{equation}\label{eq:WExpanded} 
    \sqrt{\tfrac{\pi}{2}} \cdot W(z^\alpha,z_{\mathrm{cf}},\tau)=W_{\mathrm{bulk}}(z^\alpha,\tau)+z_{\mathrm{cf}} W^{(1)}(z^\alpha,z_{\mathrm{cf}},\tau)+\mathcal{O}(z_{\mathrm{cf}}^2)\, ,
\end{equation}
with
\begin{align}\label{eq:Wbulk} 
    W_{\mathrm{bulk}}(z^\alpha,\tau)=&\frac{1}{2}M^a \widetilde{\kappa}_{a \beta\gamma}z^\beta z^\gamma-\tau K_\alpha z^\alpha+\left(P_\beta- M^a a_{a\beta}\right)z^\beta+\left(P_0-\frac{1}{24}M^a \tilde{c}'_a\right)  \nonumber \\
    &-\frac{1}{(2\pi)^2}\sum_{\tilde{\mathbf{q}}\neq \tilde{\mathbf{q}}_{\mathrm{cf}}} \mathscr{N}_{\tilde{\mathbf{q}}}\,\tilde{\mathbf{q}}_a M^a\,\text{Li}_2(\mathrm{e}^{2\pi \I \tilde{\mathbf{q}}_\alpha z^\alpha})\,  ,
\end{align}
where we introduced
\begin{equation}
\tilde{c}'_a\coloneqq \tilde{c}_a+n_{\mathrm{cf}} \delta_{a,1}\, .
\end{equation}

To achieve a small value for $W$ in the vacuum,
we want to choose the remaining fluxes such that $W_{\mathrm{bulk}}$ becomes small.
This is because $z_{\mathrm{cf}} W^{(1)}(z^\alpha,z_{\mathrm{cf}},\tau)$ is typically exponentially as we demonstrate further below.
Following \cite{Demirtas:2020ffz},
the choice
\begin{equation}
P_\beta= M^a a_{a\beta} \kom P_0=\frac{1}{24}M^a \tilde{c}'_a
\end{equation}
is convenient since it enforces a cancellation in the polynomial part of the bulk superpotential $W_{\mathrm{bulk}}$, thereby making it homogeneous and quadratic in $(z^\alpha,\tau)$ \cite{Demirtas:2019sip}.

Next, at linear order in $z_{\mathrm{cf}}$, we have
\begin{align}
    W^{(1)}(z^\alpha,z_{\mathrm{cf}},\tau)=&-M\frac{n_{\mathrm{cf}}}{2\pi \I }\Bigl(\log(-2\pi \I  z_{\mathrm{cf}})-1\Bigr)-\tau K+ \widetilde{\kappa}_{1a\gamma} M^az^\gamma+P_1-a_{1b}M^b \nonumber\\
    &+\frac{1}{2\pi \I  }\sum_{\mathbf{q}\neq \tilde{\mathbf{q}}_{\mathrm{cf}}}
     \tilde{\mathbf{q}}_1 (\tilde{\mathbf{q}}_a\,M^a)\, \mathscr{N}_{\tilde{\mathbf{q}}}\,\text{Li}_1(\mathrm{e}^{2\pi \I  \tilde{\mathbf{q}}_\alpha z^\alpha})\, ,
\end{align}
where we have defined
\begin{equation}\label{eq:coniMdef}    
M\coloneqq  M^1\, ,\; K\coloneqq K_1
\end{equation} 
and the number of conifolds is given by $n_{\mathrm{cf}}=\mathscr{N}_{\tilde{\mathbf{q}}_{\text{cf}}}$.
The $F$-flatness condition for the conifold modulus $z_{\text{cf}}$ derived from this leading-order superpotential is satisfied for
\begin{equation}
\label{eq:conifold_vev}
    \langle| z_{\mathrm{cf}}|\rangle= \frac{1}{2\pi}\exp\Biggl(-\frac{2\pi K'}{(g_s M) \, n_{\mathrm{cf}}}\Biggr)\, ,
\end{equation}
where we introduced
\begin{equation}\label{eq:Kprime} 
 K' = K-g_{s}\widetilde{\kappa}_{1a\beta}M^{a} \text{Im}(z^{\beta}) \, ,
\end{equation}
neglecting terms of the order $\mathrm{e}^{2\pi \I  \tilde{\mathbf{q}}_\alpha z^\alpha}$.
Importantly,
provided $K'/M>0$,
$z_{\text{cf}}$ is stabilised at exponentially small values giving rise to a warped throat region. 
Note that $Q_{\text{flux}}^{\text{throat}}={K'}{M}>0$ is a measure for the D3-charge from fluxes residing local conifold regions.

Provided that the conifold modulus $ \langle|z_{\mathrm{cf}}|\rangle$
as computed from \eqref{eq:conifold_vev} is very small,
the $F$-term conditions for the remaining fields $(z^\alpha,\tau)$ can be studied independently of $z_{\mathrm{cf}}$
by working with $W_{\text{bulk}}(z^\alpha,\tau)$ in \eqref{eq:Wbulk} \cite{Demirtas:2020ffz}, see also \cite{Alvarez-Garcia:2020pxd}.
We introduce the quantities
\begin{equation}
	N_{\alpha\beta}\coloneqq M^a \kappa_{a\alpha\beta}\kom p^\alpha\coloneqq  N^{\alpha\beta} K_\alpha\, .
\end{equation}
The conditions for the solutions proposed in \cite{Demirtas:2019sip,Demirtas:2020ffz} are
\begin{align}
    \label{eq:coniPFV}
    \det N\neq 0\, \kom\; \vec{p}\in \mathcal{K}_{\mathrm{cf}}\, \kom\; K_{\alpha}p^\alpha=0\, \kom\; a_{\alpha b}M^b\in \mathbb{Z}\, \kom\;  \tilde{c}'_a M^a\in 24\mathbb{Z}\, .
\end{align}
Provided that these conditions can be met simultaneously and we momentarily neglect the exponential terms in $W_{\text{bulk}}$ in \eqref{eq:Wbulk},
the remaining $F$-term conditions $D_{\alpha}W=D_{\tau}W=0$ are satisfied along a one-dimensional locus given by
\begin{equation}
    z^\alpha=p^\alpha \tau\, .
\end{equation}
This locus defines a so-called \emph{perturbatively flat vacuum} (PFV) \cite{Demirtas:2019sip} in the presence of a conifold which refer to as a \emph{conifold PFV}.

After integrating out the $z^{\alpha}$,
we are left with an effective theory for a single complex field $\tau$,
which we call the PFV effective theory,
specified by an effective superpotential
\begin{equation}\label{eq:WeffExp} 
    W^{\text{eff}}_{\text{bulk}}(\tau) = \sum_{N=1}^{\infty}\, W_N\; , \quad W_N \coloneqq -\frac{1}{(2\pi)^2} 
\sum_{\mathbf{p}_{\text{int}}\cdot\tilde{\mathbf{q}}=N}\mathscr{N}_{\tilde{\mathbf{q}}}\,\tilde{\mathbf{q}}_a M^a\,\text{Li}_2\Bigl(\mathrm{e}^{\frac{2\pi \I}{\mathtt{r}} N \t}\Bigr)\, .
\end{equation}
In the PFV effective theory,
a supersymmetric minimum for $\tau$ arises from the racetrack mechanism through the competition of consecutive terms $W_{N}$ with hierarchical coefficients.
This vacuum specified by $\langle \tau \rangle_{\mathrm{PFV}}$, $\langle  z^\alpha \rangle_{\mathrm{PFV}}=p^\alpha \langle \tau \rangle_{\mathrm{PFV}}$ in the PFV effective theory
serves as an initial guess to numerically find the solutions $\langle \tau \rangle_{\text{F}}, \langle z^{\alpha} \rangle_{\text{F}}, \langle z_{\text{cf}} \rangle_{\text{F}}$ to the full $F$-term conditions without making the PFV approximation.
For later purposes,
we define the VEV of the flux superpotential
\begin{equation}
 W_0\coloneqq  \langle |W_{\text{flux}}|\rangle_{\text{F}}
\equiv \Bigl|W_{\text{flux}}\Bigl(\langle \tau \rangle_{\text{F}}, \langle z^{\alpha} \rangle_{\text{F}}, \langle z_{\text{cf}} \rangle_{\text{F}}\Bigr)\Bigr|\,.
\end{equation}

\begin{figure}[!t]
\centering
\includegraphics[width=0.85\linewidth]{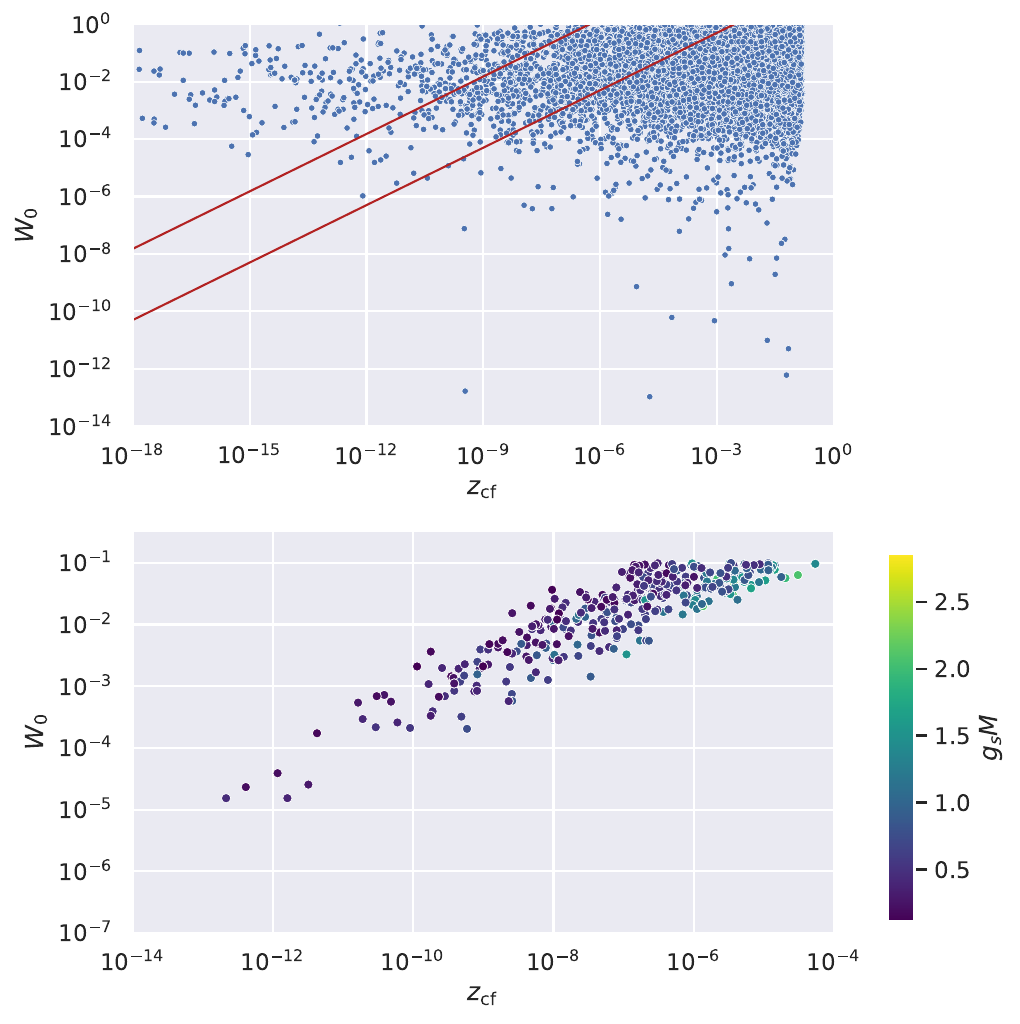}
\caption{$W_0$ as a function of the conifold modulus $z_{\text{cf}}$ for the $33{,}371$ anti-D3-brane PFVs.
In the upper plot,
the red lines indicate the alignment bounds $0.1 \le \Xi \le 10$ as defined below in Eq.~\eqref{eq:throat_tuning_maintext}.  
In the lower panel,
we zoom in on the remaining 396 points satisfying $0.1 \le \Xi \le 10$ together with $g_s<0.4$ and $W_0<0.1$. 
}  
\label{fig:W0_zcf_comb}
\end{figure}

Putting everything together, we are now ready to perform a large scale scan for conifold PFVs in the Calabi-Yau orientifolds found in \S\ref{sec:polytopeselection}.
A detailed algorithm to find such flux configurations was described in \cite{McAllister:2024lnt} to which we refer for further details.\footnote{See \cite{Dubey:2023dvu,Ebelt:2023clh,Krippendorf:2023idy,Chauhan:2025rdj} for alternative algorithmic strategies to construct flux vacua.}
By employing this algorithm, we have constructed 240{,}480{,}253 flux choices across the 416 Calabi-Yau orientifolds found in \S\ref{sec:polytopeselection}.
In these,
we identified 33{,}371 conifold PFVs with $M>12$ and $-\vec{M}\cdot\vec{K}=Q_O+2$ which we call \emph{anti-D3-brane PFVs}, see Fig.~\ref{fig:W0_zcf_comb}.
Out of these, $396$ flux vacua are in a parameter regime suitable for the anti-D3 uplift as we will make more precise in the subsequent section, see in particular Eq.~\eqref{eq:throat_tuning_maintext}.

\subsection{Kähler moduli stabilisation and uplift}\label{ss:modstab}

Thus far, we found a supersymmetric minimum for the complex structure moduli and the axio-dilaton from fluxes.
Provided that the values for the parameters $W_0$, $g_s$ and $\langle z_{\mathrm{cf}}\rangle_{\text{F}}$ are small,
we can stabilise the remaining Kähler moduli
following \cite{Kachru:2003aw}.
That is,
we first find a supersymmetric AdS vacuum at a point $T_{\text{AdS}}$ in the Kähler moduli space without the effects from the anti-D3-brane by solving $D_i W = \partial_i W + \calK_i W =0$ for all Kähler moduli $T_{i}$ where the Kähler potential $\mathcal{K}$ is given by \eqref{eq:detailedform2}. 
To this end, we evaluate the superpotential \eqref{eq:wlo3} using the VEVs for $z^{a}$ and $\tau$ such that
\begin{equation}\label{eq:wlo3evaluated}
    W = W_0 + \sqrt{\frac{2}{\pi}}\frac{1}{4\pi^2}\, \sum_D \mathrm{e}^{-2\pi T_D/c_D}\, .
\end{equation}
We then proceed by first finding a \textit{KKLT point} $t^i\in \calK_X$ in Kähler moduli space where \cite{Demirtas:2021nlu}
\begin{equation}
    \label{eq:T0def} \Re{T_i} = \Re{T_i^0} \coloneqq  \frac{c_i}{2\pi}\log{W_0^{-1}}\, .
\end{equation}
Even though the $F$-terms $D_iW$ do not vanish at such KKLT points,
they serve as good initial guesses for a numerical search for points satisfying $D_i W=0$.
Indeed, starting from $T^0_i$,
we use Newton's method to find the VEVs $T_i = \langle T_i \rangle_{\mathrm{F}}$ in Kähler moduli space where $D_iW=0$.
Such a solution corresponds to a supersymmetric AdS$_4$ vacuum with all moduli stabilised.
Contrary to \cite{Demirtas:2021nlu}, our examples also possess Klebanov-Strassler throats.

At this point, let us make one conceptual remark.
A widespread confusion about the KKLT scenario is the belief that one first needs to construct a fully consistent, physical, supersymmetric AdS$_4$ vacuum, and then simply adds an anti-D3-brane to achieve uplift to a de Sitter solution. However, this is not possible due to the following simple argument: an anti-D3-brane carries a charge of $-1$ unit relative to a D3-brane, so the D3-brane tadpole condition \eqref{eq:gaussLaw} differs by one unit between the two configurations, and thus cannot be precisely satisfied in both cases.
Even so, the AdS solution --- which we shall occasionally refer to as a \emph{AdS precursor} --- remains a valuable reference point. In our examples, where the number of moduli is large and the scalar potential exhibits considerable complexity, the approximate solutions derived from the AdS VEVs provide essential starting points for the full analysis.

The final step is to include the anti-D3-brane potential $V_{\overline{D3}}$ as defined in \eqref{eq:anti-D3-potential0} and use $T_{\text{AdS}}$ as an initial guess
to numerically search for a de Sitter minimum $T_{\text{dS}}$ of the full potential \eqref{eq:Vfull}.
To avoid inducing a runaway instability through the uplifting term $V_{\overline{D3}}$ in Eq.~\eqref{eq:anti-D3-potential0},
we have to ensure that $V_{\overline{D3}}$ is of the same order as $\langle V_{F}\rangle_{\text{AdS}}=-3|W|^{2}\mathrm{e}^{\calK}$.
This leads to the constraint
\begin{equation}\label{eq:throat_tuning_maintext}
\Xi \coloneqq \dfrac{V_{\overline{D3}}}{|V_F|} \approx \frac{|z_{\text{cf}}|^{\frac{4}{3}}}{|W_0|^2}\frac{\mathcal{V}_E^{\frac{2}{3}}\widetilde{\mathcal{V}}^{\frac{1}{3}}}{(g_s M)^2}\cdot \zeta \sim 1
\,, 
\end{equation}   
where we introduced the constant $\zeta\approx 114$ \cite{McAllister:2024lnt}.
A (AdS) vacuum satisfying \eqref{eq:throat_tuning_maintext} will be called \textit{well-aligned}.
For such AdS vacua,
we attempt to find de Sitter critical points by solving
\begin{equation}
    \d_i V = 0
\end{equation}
for the full scalar potential $V$ in \eqref{eq:Vfull}.
In this last step,
the VEVs for the complex structure moduli and axio-dilaton shift due to the uplift which we carefully take into account.

 \newpage

\section{Candidate KKLT de Sitter Vacua}\label{sec:examples}

\begin{table}
    \centering
    \begin{tabular}{c|c|c}
   \textbf{Condition}      & \textbf{Number of configurations} & \textbf{Explanation} \\ \hline 
    & & \\[-1.3em]
   $3\le h^{2,1} \le 8$      &   202{,}073 polytopes  & \S\ref{sec:polytopeselection} \\[0.2em]
   trilayer, $\Delta$ and $\Delta^\circ$ favorable     &  3187 polytopes   & \S\ref{sec:polytopeselection} \\[0.2em]  
   $h^{1,1}$ cuts and $\ge h^{1,1}$ rigid divisors  & 322   polytopes   & \S\ref{sec:polytopeselection} \\[0.2em]
   conifold consistent with KKLT point  &  416 conifolds  & \S\ref{sec:polytopeselection} \\[0.2em]
   fluxes giving conifold PFV   &  240{,}480{,}253 conifold PFVs  & \S\ref{sec:fluxes} \\[0.2em]
   $M>12$; one anti-D3-brane     & 33{,}371  anti-D3-brane PFVs  & \S\ref{sec:fluxes}\\[0.2em]
   de Sitter vacuum & 30 de Sitter vacua & \S\ref{sec:examples}, Tab.~\ref{tab:summary}
    \end{tabular}
    \caption{Number of configurations identified at each stage of the selection procedure. The criteria are applied cumulatively: each row reflects configurations that satisfy all conditions listed in the rows above. We do not enumerate the exponentially large number of inequivalent triangulations of $\Delta^{\circ}$, though our analysis permits the exploration of any phase of the extended Kähler cone of $\Delta^{\circ}$. As for flux configurations leading to PFVs, the scan was broad in scope, though not fully comprehensive.}
    \label{tab:scan_summary}
\end{table}

Let us now describe the
de Sitter vacua 
obtained in the leading-order EFT \S\ref{sec:EFTs} via a large-scale computational search employing the methods described in \S\ref{sec:methods}.
We collected the results for the various steps laid out in \S\ref{sec:methods} in Tab.~\ref{tab:scan_summary}.
As summarised in Tab.~\ref{tab:summary},
we found five distinct compactifications with a total of 30 such vacua.
We will describe two of these examples below.
Our data and demo notebooks to validate our solutions are publicly available on \href{https://github.com/AndreasSchachner/kklt_de_sitter_vacua}{{GitHub}}.

\subsection{Example 1: $h^{1,1}=150$, $h^{2,1}=8$} \label{sec:manwe} 
 
First,
we study the polytope $\De$ specified by the vertices 
\begin{equation}\label{eq:manwe_poly}
    \left(\begin{array}{ccccccccc}
    \phantom{-}1& -1& -1& -1& -1& -1& -1& -1& -1\\
    -1&  \phantom{-}2& -1& -1& -1&  \phantom{-}0& -1&  \phantom{-}0&  \phantom{-}0\\ 
    -1&  \phantom{-}1&  \phantom{-}0&  \phantom{-}2&  \phantom{-}2&  \phantom{-}0&  \phantom{-}0&  \phantom{-}0&  \phantom{-}1\\  
    -1&  \phantom{-}1&  \phantom{-}0&  \phantom{-}0&  \phantom{-}1&  \phantom{-}2&  \phantom{-}2&  \phantom{-}1&  \phantom{-}0\\
    \end{array}\right)\,.
\end{equation}
An FRST of $\De$ and of its polar dual $\De^{\circ}$ give rise to a mirror pair of smooth Calabi-Yau threefolds $(\Xt,X)$ with Hodge numbers $h^{1,1}(\Xt)=h^{2,1}(X)=8$ and  $h^{2,1}(\Xt)=h^{1,1}(X)=150$.
In a suitable FRST $\widetilde{\mathscr{T}}$ of $\De$, there exists a conifold curve with  $n_{\text{cf}}=2$ allowing us to engineer a conifold singularity in complex structure moduli space of $\Xt$ as described in \S\ref{sec:EFTs}.
Using the methods of \cite{Moritz:2023jdb},
one easily verifies that there exists a sign-flip orientifold with $h^{1,1}_-(X/\mathcal{I})=h^{2,1}_+(X/\mathcal{I})=0$ for any such $X$ defined by an FRST of $\Dec$.
There exist $152$ pure rigid prime toric divisors
of which $35$ host O7-planes that support $\mathfrak{so}(8)$ stacks.

A conifold PFV is furnished by the following choice of fluxes in $H^{3}(X,\bbZ)$ (cf.~\S\ref{sec:fluxes})
\begin{align}
    \vec{M} &= \begin{pmatrix} 16& 10& -26& 8& 32& 30& 18& 28  \end{pmatrix}^\top\, , \\[0.4em]
    \vec{K}&=\begin{pmatrix}-6& -1& 0& 1& -3& 2& 0& -1\end{pmatrix}^\top\,,
\end{align}
These flux vectors satisfy $-\vec{M}\cdot\vec{K}  =162 $ which implies that the presence of a single anti-D3-brane is required to satisfy Gauss's law \eqref{eq:gaussLaw}.
The PFV locus can be found from the leading terms in the effective superpotential \eqref{eq:WeffExp}
which are here given by
\begin{equation}\label{eq:manwew}
    W_{\text{PFV}} = \frac{1}{\sqrt{8\pi^5}}\biggl(14\,\mathrm{e}^{2\pi \I  \t \cdot \frac{1}{40}} - 80\,\mathrm{e}^{2\pi \I \t \cdot \frac{2}{40}} +
    118\,\mathrm{e}^{2\pi \I \t \cdot \frac{3}{40}} +\ldots
    \biggr)\,.
\end{equation} 
Numerical minimisation of $D_{z^a}W_{\text{flux}}$ and $D_{\tau}W_{\text{flux}}$
with the PFV minimum as a starting guess allows us to find the true $F$-term solutions of the full flux superpotential \eqref{eq:WExpanded}.
This leads to
\begin{equation}
    g_s = 0.0732\, , \ z_{\text{cf}} = 1.390\times10^{-7}\, , \ W_0 = 0.0103\, , \ \text{and}\ g_sM = 1.171\,.
\end{equation}

\begin{figure}[!t]
\centering 
\includegraphics[width=.8\linewidth]{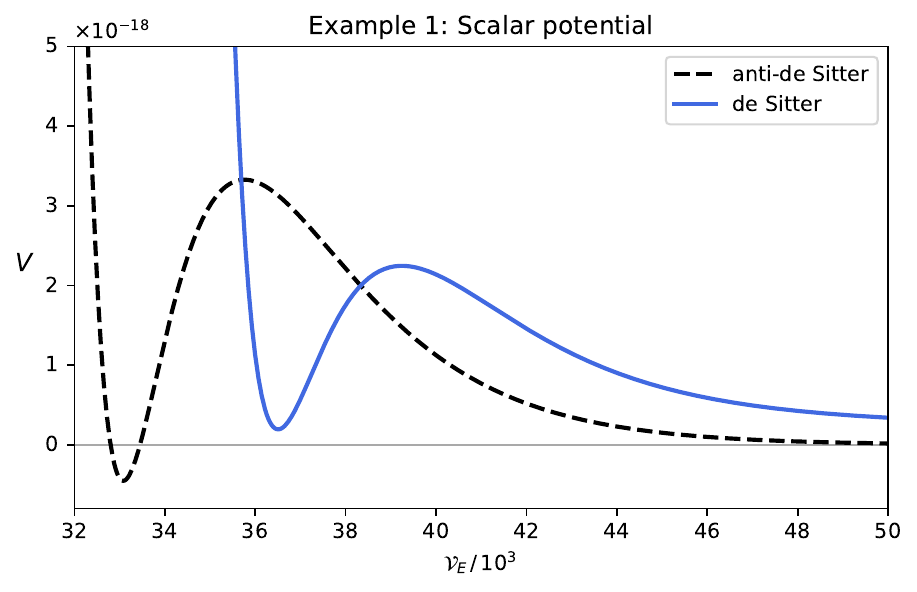}
\caption{Kähler moduli potential before and after uplift for Example 1 in \S\ref{sec:manwe}.}
\label{fig:manwe_uplift}
\end{figure}

Following the procedure described in \S\ref{sec:methods}, we next stabilise the Kähler moduli.
The non-perturbative superpotential \eqref{eq:wlo3} receives $h^{1,1}(X)+1=151$ contributions normalising the Pfaffians as in \eqref{eq:K0} with $n_{D}=1$: there are $35$ terms from gaugino condensation on $\mathfrak{so}(8)$ stacks with  $c_D=6$ and $116$ terms from Euclidean D3-branes with $c_D=1$.
We neglect the contribution from a single Euclidean D3-brane intersecting the conifold, cf.~\S\ref{sec:polytopeselection}.
Using the Kähler potential in \eqref{eq:detailedform2} and holomorphic Kähler coordinates \eqref{eq:detailedform4},
we initially find a candidate AdS minimum $T_{\text{AdS}}$ for the Kähler moduli inside the torically extended Kähler cone of $\Dec$ shown as the black line in Fig.~\ref{fig:manwe_uplift}.
At the point $T_{\text{AdS}}$, the $\a'$-corrected string-frame volume $\calV$ and Einstein-frame volume $\calV_{\text{E}}$ are
\begin{equation}
     \calV = \calV^{(0)} + \de\calV_{\alpha'^3} + \de\calV_{\text{WSI}} = 665.45 - 0.34  -0.58 = 664.53\; ,\quad  \calV_{\text{E}} = 3.30 \times 10^4\,.  
\end{equation}

\begin{figure}[!t]
\centering
\includegraphics[width=0.8\linewidth]{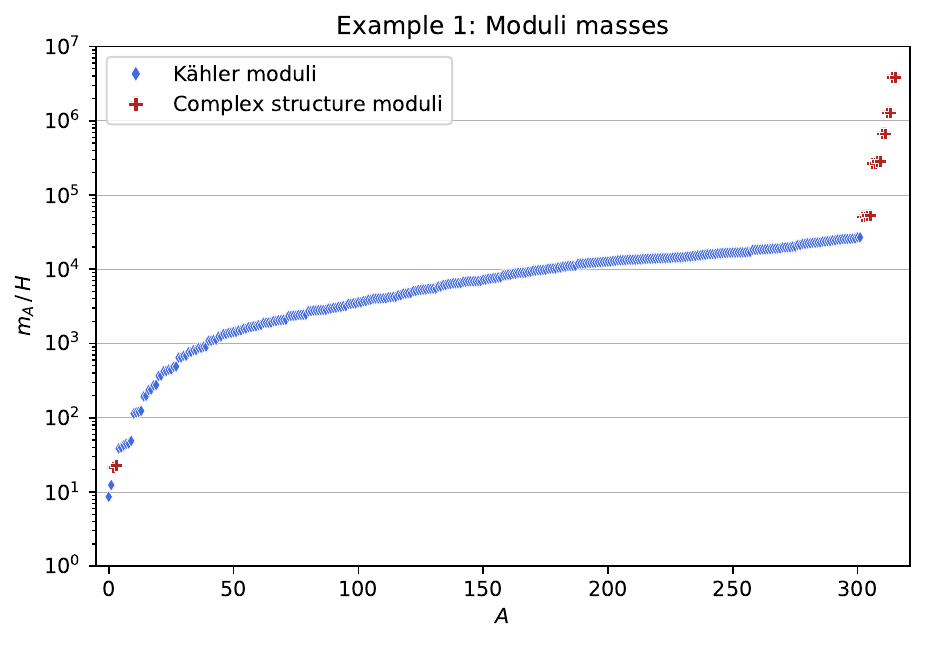}
\caption{Plot of the moduli masses in the de Sitter vacuum discussed in \S\ref{sec:manwe}.} 
\label{fig:manwe_masses}
\end{figure}

Next, we incorporate supersymmetry breaking by adding the anti-D3 brane potential \eqref{eq:anti-D3-potential0}
and iteratively uplift the Kähler and complex structure moduli as discussed in \S\ref{ss:modstab}.
This process leads to a candidate de Sitter vacuum at $T_{\text{dS}}$ in Kähler moduli space shown in Fig.~\ref{fig:manwe_uplift} with
\begin{equation}
    V_{\text{dS}} = +1.937\times10^{-19} M_\text{pl}^4\,,
\end{equation}
and the complex structure parameters
\begin{align}
    g_s &= 0.0657\,, \; W_0 = 0.0115\,, \; z_{\text{cf}} = 2.822\times10^{-8}\,, \; g_sM = 1.051 \,. \label{eq:param4uplift}
\end{align}
The $\a'$-corrected string-frame and Einstein-frame volumes are given by
\begin{equation}
     \calV  = \calV^{(0)} + \de\calV_{\alpha'^3} + \de\calV_{\text{WSI}} = 614.83 - 0.34  -0.58 = 613.91\; , \quad  \calV_{\text{E}} \approx 3.65 \times 10^4\,. 
\end{equation}
The mass spectra of the moduli is shown in Fig.~\ref{fig:manwe_masses} in terms of the Hubble scale,
\begin{equation}
    H_{\text{dS}} = \sqrt{\tfrac{1}{3}V_{\text{dS}}} = 2.5 \times 10^{-10} M_\text{pl}\,.
\end{equation}
The lightest mode, corresponding to a Kähler modulus, has mass
\begin{equation}
    m_{\text{min}} = 8.616 H_{\text{dS}}\,.
\end{equation}

\begin{figure}[!t]
\centering
\includegraphics[width=0.8\linewidth]{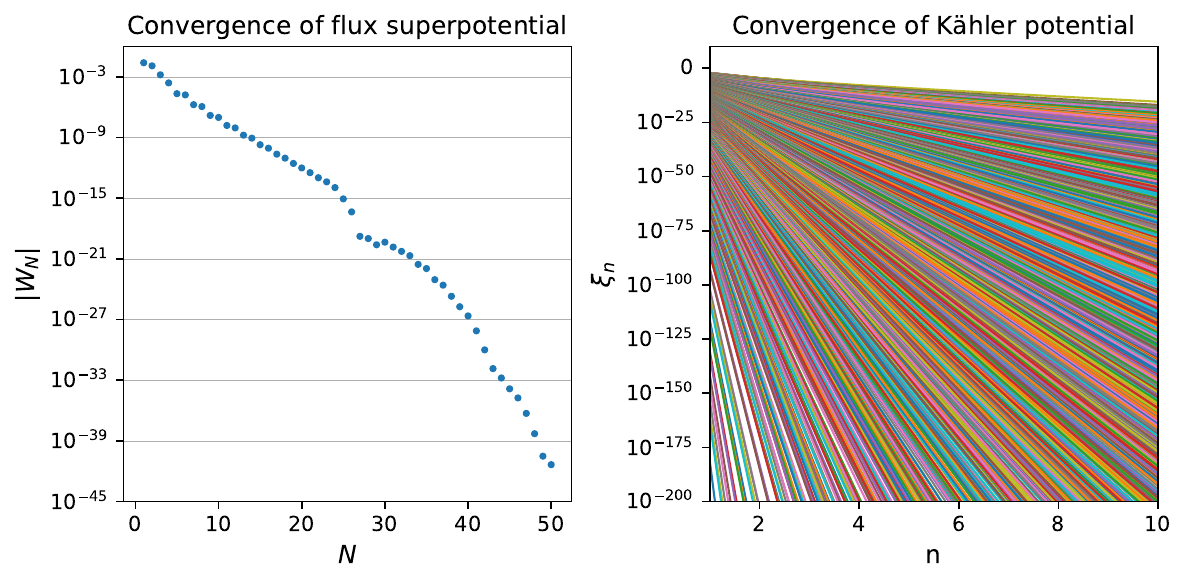}
\caption{Convergence checks for Example $1$ in \S\ref{sec:manwe}. \emph{Left:} 
The various superpotential terms \eqref{eq:WeffExp} converge for the bulk complex structure moduli $z^{\alpha}$ in the dS minimum  
\emph{Right:} Worldsheet instanton contributions from a random sample of
$2{,}643$ potent rays spanning a $144$-dimensional sub-cone of the Mori cone $\mathcal{M}_{X}$ of $X$.
}\label{fig:manwe_W_rainbow}
\end{figure}

Let us check that the truncations made in the above minimum are consistent.
To begin with,
we recall from \eqref{eq:WeffExp} that the flux superpotential effectively corresponds to a sum over mirror worldsheet instantons.
Convergence can be easily demonstrated by plotting  the individual summands $|W_{N}|$ defined in Eq.~\eqref{eq:WeffExp} which is shown on the left-hand side of Fig.~\ref{fig:manwe_W_rainbow}.
Similarly, to validate the convergence of worldsheet instantons contributing to the Kähler potential \eqref{eq:detailedform2} and the Kähler moduli \eqref{eq:detailedform4},
we compiled a sample of $2{,}643$ random potent rays inside low-dimensional faces of the Mori cone $\mathcal{M}_{X}$ of $X$.
For each such ray $\{n\mathcal{C}\,|\, n\in\mathbb{Z}_{+}\}$ from a curve $\mathcal{C} \in \mathcal{M}_{X}$ with $\mathbf{q}\in \mathcal{M}_{X}\cap H_2(X,\mathbb{Z})$,
we calculate the quantity
\begin{equation}\label{eq:xin}
    \xi_n(\mathbf{t},\mathcal{C}) \coloneqq \bigl |\mathscr{N}_{n\mathbf{q}}\, \mathrm{e}^{-2\pi n\, \mathbf{q}\cdot\mathbf{t}}\bigl |\, .
\end{equation}
We plot the result on the right-hand side of Fig.~\ref{fig:manwe_W_rainbow} which demonstrates that the corrections \eqref{eq:xin} are exponentially decaying with $n$.

\subsection{Example 4: $h^{1,1}=93$, $h^{2,1}=5$} \label{sec:aule}

Next,
we study the polytope $\De$ specified by the vertices 
\begin{equation}\label{eq:aule_poly}
    \left(\begin{array}{ccccccccc}
    \phantom{-}1& -1& -1& -1& -1& -1& -1& -1& -1\\
    -1&  \phantom{-}1& \phantom{-}0& \phantom{-}0& \phantom{-}0&  \phantom{-}1& \phantom{-}1&  \phantom{-}2&  \phantom{-}1\\ 
    -1&  \phantom{-}1&  \phantom{-}0&  \phantom{-}0&  \phantom{-}1&  \phantom{-}0&  \phantom{-}2&  \phantom{-}1&  \phantom{-}1\\  
    -1&  \phantom{-}0&  \phantom{-}0&  \phantom{-}1&  \phantom{-}0&  \phantom{-}1&  \phantom{-}1&  \phantom{-}2&  \phantom{-}2\\
    \end{array}\right)\,.
\end{equation}
An FRST of $\De$ and of its polar dual $\De^{\circ}$ give rise to a mirror pair of smooth Calabi-Yau threefolds $(\Xt,X)$ with Hodge numbers $h^{1,1}(\Xt)=h^{2,1}(X)=5$ and  $h^{2,1}(\Xt)=h^{1,1}(X)=93$.
In a suitable FRST $\widetilde{\mathscr{T}}$ of $\De$, there exists a conifold curve with  $n_{\text{cf}}=2$ allowing us to engineer a conifold singularity in complex structure moduli space of $\Xt$ as described in \S\ref{sec:EFTs}.
Using the methods of \cite{Moritz:2023jdb},
one easily verifies that there exists a sign-flip orientifold with $h^{1,1}_-(X/\mathcal{I})=h^{2,1}_+(X/\mathcal{I})=0$ for any such $X$ defined by an FRST of $\Dec$.
There exist $96$ pure rigid prime toric divisors
of which $21$ host O7-planes that support $\mathfrak{so}(8)$ stacks.

In the relevant FRST $\widetilde{\mathscr{T}}$ of $\De$  specified in the ancillary files in the \href{https://github.com/AndreasSchachner/kklt_de_sitter_vacua}{GitHub repository},
we found a conifold PFV given by the vectors
\begin{align}\label{eq:FluxesAule} 
    \vec{M} &= \begin{pmatrix}  20&   4&   8& -18& -20  \end{pmatrix}^\top\, , \\[0.4em]
    \vec{K} &=\begin{pmatrix}-5& -1&  0&  1& -1\end{pmatrix}^\top\,,
\end{align}
with $-\vec{M}\cdot\vec{K}=102$ necessitating a single anti-D3-brane to satisfy Gauss's law \eqref{eq:gaussLaw}.
In this case,
we obtain a solution to the $F$-term conditions $D_{z^{a}}W=D_{\tau}W=0$ with parameters 
\begin{equation}
    g_s = 0.0410\, , \ z_{\text{cf}} = 2.369\times10^{-6}\, , \ W_0 = 0.0525\, , \ \text{and}\ g_sM = 0.821\,.
\end{equation}

\begin{figure}[!t]
\centering
\includegraphics[width=0.85\linewidth]{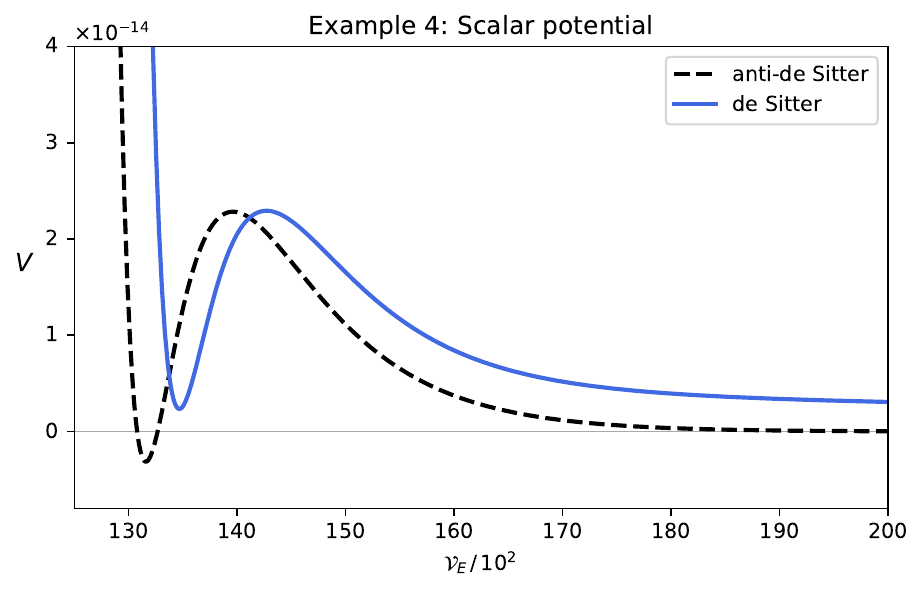}
\caption{Kähler moduli potential before and after uplift for Example 4 in \S\ref{sec:aule}.}\label{fig:aule_uplift}
\end{figure}

\begin{figure}[!t]
\centering
\includegraphics[width=0.85\linewidth]{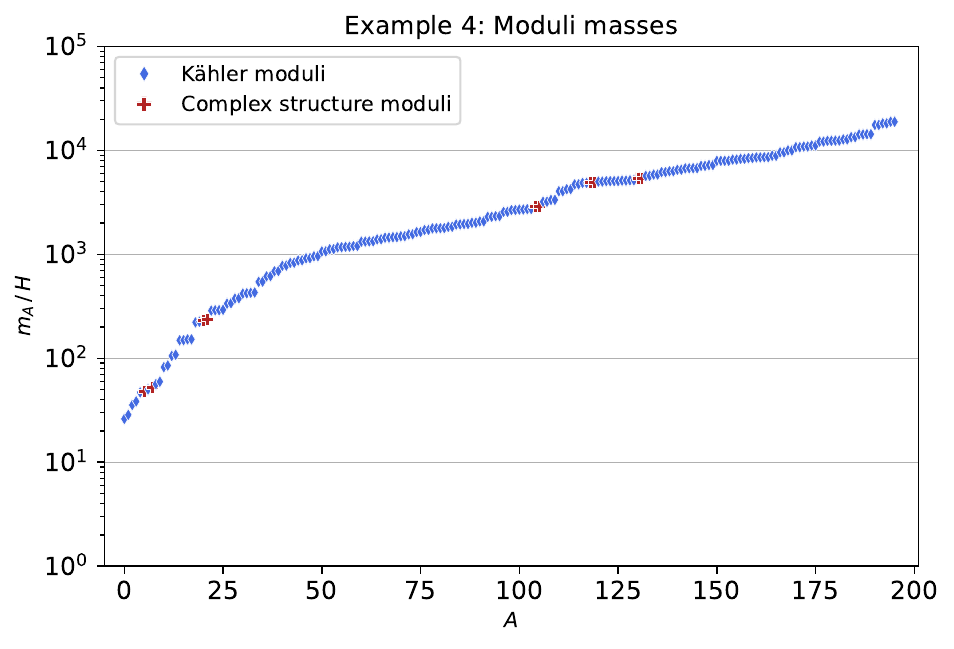}
\caption{Plot of the moduli masses in the de Sitter vacuum discussed in \S\ref{sec:aule}.}\label{fig:aule_masses}
\end{figure}

In contrast to the previous example in \S\ref{sec:manwe}, 
we find 36 KKLT points in this compactification which leads to 29 AdS precursors. 
Focussing on a single AdS precursor from this list, we compute the corrected Einstein frame volume in the minimum
\begin{equation}
    \calV_E = 1.310 \times 10^4\,.   
\end{equation}
Including the anti-D3-brane, 
we find a de Sitter vacuum with vacuum energy 
\begin{equation}
    V_{\text{dS}} = +2.341\times10^{-15}\, M_{\text{pl}}^4\,.
\end{equation}
The parameters in the non-supersymmetric minimum are given by
\begin{equation}
    g_s = 0.0404\, , \ z_{\text{cf}} = 1.965\times10^{-6}\, , \ W_0 = 0.0539\, , \ \text{and } \ g_sM = 0.808\,.
\end{equation}
At the minimum $T_{\text{dS}}$ for the Kähler moduli,
the fully corrected Einstein frame volume is given by
\begin{equation}
    \calV_E = 1.340 \times 10^4\,. 
\end{equation}
The scalar potential for the Kähler moduli before and after uplift is shown in Fig.~\ref{fig:aule_uplift}.
As shown in Fig.~\ref{fig:aule_masses}, the vacuum is free of tachyons with the smallest modulus mass corresponding to
\begin{equation}
    m_{\text{min}} = 26.157\, H_{\text{dS}}\,.
\end{equation}
For the convergence tests, we refer to \cite{McAllister:2024lnt}.

\begin{figure}[!t]
\centering
\includegraphics[width=0.85\linewidth]{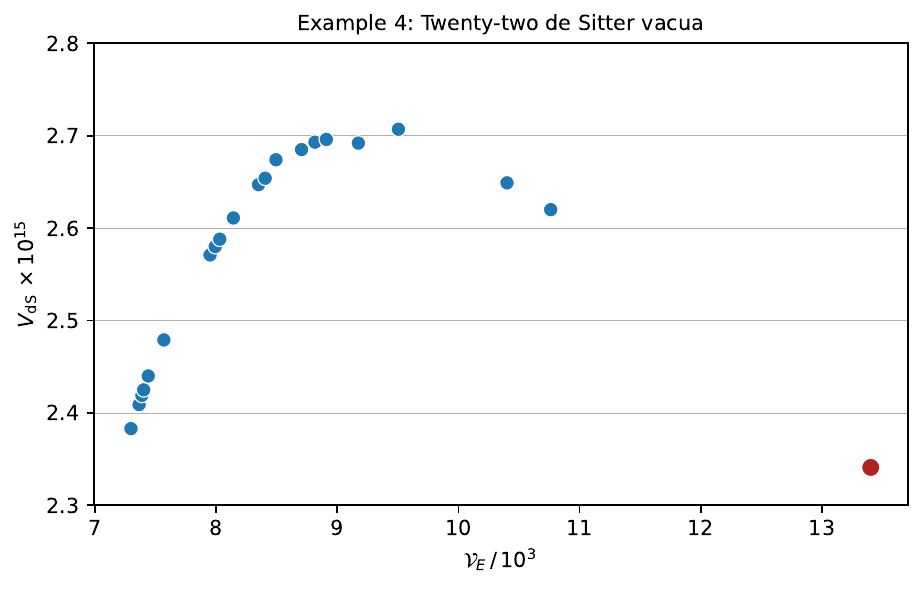}
\caption{For the same choice of flux quanta \eqref{eq:FluxesAule},  we found 21 additional de Sitter vacua (blue dots) in the extended Kähler cone.
The de Sitter solution discussed in the main text corresponds to the red dot.} \label{fig:aule_others}
\end{figure}

In addition to the de Sitter vacuum presented above,
there are 28 additional AdS precursors of which 21 lead to additional de Sitter vacua.
The vacuum energy and Einstein frame volumes for these solutions are shown in Fig.~\ref{fig:aule_others}.

\FloatBarrier

\newpage

\vphantom{10cm}

\vfill

\FloatBarrier

\newpage

\section{Conclusions}\label{sec:conclusion}

%This work has laid the foundations for a systematic study of KKLT de Sitter vacua.
In our search for de Sitter vacua of string theory,  
we obtained 100 million type IIB flux compactifications on Calabi-Yau orientifolds
out of which 33{,}371 featured a metastable anti-D3-brane and a Klebanov-Strassler throat \cite{Kachru:2002gs}.
Ultimately, following \cite{Kachru:2003aw}, we isolated five compactifications exhibiting local minima of the moduli potential with positive vacuum energy.
They constitute \emph{candidate} de Sitter vacua of string theory, insofar as the effective theory in which our analysis was performed uses certain approximations: we worked at string tree level but to all orders in $\alpha'$ in the closed string sector, whilst we retained only the leading order in both expansions for the anti-D3-brane potential \cite{Kachru:2002gs}.

The vacuum structure in our models is governed by non-perturbative effects: worldsheet instanton corrections to the Kähler potential as well as Euclidean D3-brane contributions to the superpotential.
Here, only the Pfaffian numbers remained as the only undetermined parameters in the leading-order effective theory.
Significant advances in several areas are prerequisites to decide whether the solutions presented here correspond to genuine de Sitter vacua of string theory.
In particular, it will be necessary to compute the corresponding warped metric, string loop corrections to the Kähler coordinates and the Kähler potential in Calabi-Yau orientifolds \cite{Berg:2005ja,Kim:2023sfs,Kim:2023eut},
and $\alpha'$ corrections to the potential of anti-D3-branes in warped throats \cite{Junghans:2022exo,Junghans:2022kxg,Hebecker:2022zme,Schreyer:2022len,Schreyer:2024pml}, see e.g. \cite{Cho:2023mhw} for promising directions using string field theory.

Even in the absence of major conceptual breakthroughs in string theory,
meaningful progress can still be made through large-scale computational methods.  
In this initial study, we restricted our scan to cases with $h^{2,1} \leq 8$, a choice motivated by the need to keep computational demands tractable.  
However, this limitation excludes the vast majority of the landscape of flux vacua associated with toric Calabi–Yau  hypersurfaces.  
The solutions identified here thus represent only a preliminary exploration of what is likely to be a vastly richer structure within the Kreuzer-Skarke dataset.  
In particular, we note that numerous polytopes within the Kreuzer-Skarke list exhibit significantly larger values of $Q_{\text{O}}$ than those considered in our analysis, and we anticipate that these will offer especially promising directions for further study.
That said, the computational undertaking involved in this work was already considerable: the construction of over 100 million flux vacua required approximately 50 core-years of processing time.  
Extending the search to models with larger $h^{2,1}$ would demand both the creation of new algorithmic strategies like the ones described in \cite{Dubey:2023dvu,Chauhan:2025rdj} and access to computational resources well beyond the capacity of modest-scale clusters.

The compactifications developed in this work are intended to serve as a concrete testing ground for exploring the numerous outstanding questions in the field. We envisage that these constructions will provide a useful foundation for advancing the understanding of vacuum structures arising from string theory.

\section*{Acknowledgements}

I am grateful to Liam McAllister, Jakob Moritz, and Richard Nally for collaboration on this project.
I also thank Michele Cicoli, Arthur Hebecker, Sven Krippendorf, Dieter Lüst, Fernando Quevedo, Simon Schreyer and Alexander Westphal for interesting discussions.

\hfill

\newpage

\bibliographystyle{utphys}
\bibliography{Literatur}

\end{document}